\documentclass{article}
\usepackage{amsfonts}
\usepackage{amssymb}
\usepackage{amsmath}

\newenvironment{proof}[1][Proof]{\noindent\textbf{#1.} }{\ \rule{0.5em}{0.5em}}
\input{tcilatex}

\begin{document}

\begin{center}
{\LARGE On the Explanation for Quantum Statistics}

\medskip

{\Large Simon Saunders\bigskip }

\bigskip
\end{center}

\begin{quotation}
\noindent \textbf{Abstract} The concept of classical indistinguishability is
analyzed and defended against a number of well-known criticisms, with
particular attention to the Gibbs' paradox. Granted that it is as much at
home in classical as in quantum statistical mechanics, the question arises
as to why indistinguishability, in quantum mechanics but not in classical
mechanics, forces a change in statistics. The answer, illustrated with
simple examples, is that the equilibrium measure on classical phase space is
continuous, whilst on Hilbert space it is discrete. The relevance of names,
or equivalently, properties stable in time that can be used as names, is
also discussed.

\end{quotation}

\bigskip

\noindent Einstein's contributions to quantum theory, early and late, turned
on investigations in statistics - most famously with his introduction of the
light quantum, now in its second century, but equally with his penultimate
contribution to the new mechanics (on Bose-Einstein statistics) in 1924.
Shortly after, his statistics was incorporated into to the new (matrix and
wave) mechanics. But there remained puzzles, even setting to one side the
question of his last contribution (on the completeness of quantum
mechanics). A number of these centre on the concept of particle
indistinguishability, which will occupy us greatly in what follows.

To keep the discussion within reasonable bounds, and for the sake of
historical transparency, I shall use only the simplest examples, and the
elementary combinatoric arguments widely used at the time. For similar
reasons, I shall largely focus on Bose-Einstein statistics (and I shall
neglect parastatistics entirely). Hence, whilst not a study of the history
of quantum statistics, I shall be speaking to Einstein's time.

But if not anachronistic, my way of putting things is certainly
idiosyncractic, and calls for some stage-setting.

\section{The Puzzle}

These are the puzzling features to be explained: distinguishable particles,
classically, obey Maxwell-Boltzmann statistics, but so do \textit{%
indistinguishable} (permutable) particles. In quantum mechanics,
distinguishable particles also obey Maxwell-Boltzmann statistics; but 
\textit{not\ }so indistinguishable ones. There is evidently something about
the combination of permutation symmetry and quantum mechanics that leads to
a difference in statistics. What, precisely?

Were the concept of indistinguishability unintelligible from a classical
perspective, the puzzle would hardly arise, or not in this form;
indistinguishability in itself, in that case an inherently quantum concept,
would be the obvious culprit. And, indeed, the notion of classical
permutability has for the main part been viewed with suspicion (because
already incompatible with classical principles, philosophical or physical,
almost always unstated). After all, classical particles can surely be
distinguished by their trajectories (an argument I shall discuss at length).
The very concept of particle indistinguishability only came to prominence
through investigations in quantum statistics (Ehrenfest (1911) and Natansen
(1911)). It was natural to view particle indistinguishabililty as an
intrinsically quantum mechanical concept.\footnote{%
Although Schr\"{o}dinger had by the close of this period (in his last paper
prior to the series on wave mechanics) shown how one can dispense with it,
taking as distinguishable objects the modes of a system of waves as
representing the states of a gas (he was of course drawing at this point
heaviliy on de Broglie's ideas). It was an advantage of the approach,
thought Schr\"{o}dinger, that the states of the gas thus conceived obeyed
Maxwell-Boltzman statistics (see Schr\"{o}dinger (1924) and, for further
discussion, Dieks (1990); I shall come back to this point at the end).}

And yet the same concept is important to another puzzle that arises already
in classical mechanics - the \textit{Gibbs' paradox}. In particular it
explains the subtraction of a term $k\ln N!$ from the classical (Boltzmann)
entropy for $N$ identical particles\footnote{%
Meaning that differences between them, if any, are irrelevant to the
dynamics.} (to give the entropy as an extensive function of state); so much
is required if there is to be zero entropy of mixing of two samples of the
same gas. Division of the classical phase space volume by $N!$, as follows
if identical classical particles are permutable (so that phase space points
related by a permutation are identified), supplies the needed correction. It
was explained in just this way (using his own definition of the entropy
function) by Gibbs (1902 p.206-07).

But Gibbs' view of the matter found few supporters, and was rapidly
overtaken by events; so much so, that by mid-century, in reply to a related
question raised by Schr\"{o}dinger, it was not even judged worthy of mention:

\begin{quotation}
In conclusion, it should be emphasized that in the foregoing remarks
classical statistics is considered in principle as a part of classical
mechanics which deals with individuals (Boltzmann). The conception of atoms
as particles losing their identity cannot be introduced into the classical
theory without contradiction. (Stern 1949).
\end{quotation}

\noindent (Stern did not say in what the contradiction consists).

For a text book still in wide use:

\begin{quotation}
\noindent It is not possible to understand classically why we must divide by
N! to obtain the correct \ counting of states (Huang, 1963 p.154).
\end{quotation}

\noindent (although classical permutability implies it directly).

For a statement by Schr\"{o}dinger on the subject::

\begin{quotation}
It was a famous paradox pointed out for the first time by W. Gibbs, that the
same increase of entropy must not be taken into account, when the two
molecules are of the same gas, although (according to naive gas-theoretical
views) diffusion takes place then too, but unnoticeably to us, because all
the particles are alike. The modern view [of quantum mechanics] solves this
paradox by declaring that in the second case there is no real diffusion,
because exchange between like particles is not a real event - if it were, we
should have to take account of it statistically.\footnote{%
However, in quantum mechanics as classically, certainly there \textit{is}
real diffusion (of particles initially confined to a sub-volume, on removal
of a partition separating the two samples of gas, into the total volume).}
It has always been believed that Gibbs's paradox embodied profound thought.
That it was intimately linked up with something so important and entirely
new [as quantum mechanics] could hardly be foreseen. (Schr\"{o}dinger 1946
p.61).
\end{quotation}

\noindent (implicitly suggesting that the paradox could not be resolved on
purely classical grounds).

And from a recent monograph devoted to the concept of classical particle
indistinguishability:

\begin{quotation}
Prior to the description of a state by means of probability measures states
were identified with point measures. In this deterministic setting
indistinguishable objects are not conceivable. (Bach 1997, p.131)
\end{quotation}

\noindent I shall come back to Bach's treatment of classical
indistinguishability shortly.

I\ do not wish to suggest that the concept was universally rejected; Leon
Rosenfeld\footnote{%
Another exception is Hestines, who has also defended the concept of
classical indistinguishability (Hestines (1970)); but, with certain
reservations, his approach is similar to that of Rosenfeld and Bach,
discussed below, and I shall not address it separately.} spoke with approval
of Planck's appeal to indistinguishability to explain the needed subtraction
(in the 4th edition of \textit{Theorie der Warmestrahlung} of 1921),
comparing it with the discussion by Gibbs. This, thought Rosenfeld, was a
`simple and clear interpretation' but one (he went on to say) that was
rejected by Ehrenfest and Trkal (as `incomprehensible'\footnote{%
But note the word does not occur in the English-language version of their
paper.}).

However, Rosenfeldl understood Gibbs' conception of statistical mechanics as
`fundamentally idealistic' (Rosenfeld 1959 p.244); in contrast:

\begin{quotation}
\noindent The argument of Ehrenfest and Trkal is clearly inspired by the
materialist (others prefer to say `realist') attitude characteristic of
Boltzmann's thought: the problem of statistical mechanics is to give a
complete and logically coherent deduction of macroscopic laws of
thermodynamics from those of mechanics, applied to the atomic model of
matter. \textit{Considered from this point of view, it is evidently decisive}%
; and it is thus particularly significant that it does not at all impress
Planck. (Rosenfeld 1959, p.243, emphasis mine).
\end{quotation}

\noindent (Planck, Rosenfeld went on to explain, did not share Boltzmann's
`materialist' preoccupations.)

I take it to have established a \textit{prima facie} case that the concept
of classical indistinguishability, at least at a realist, microphysical
level of description, has been greatly neglected, and for the most part
dismissed out of hand. In what follows I shall first dissect with rather
more care the arguments of its critics (Section 2), and then go on to defend
the principle directly (Section 3); only if it is granted as \textit{%
classically }intelligible does the puzzle about quantum statistics arise in
the way that I\ have stated it. An answer to the puzzle follows (Section 4).
Indeed, it is not too hard to find, prefigured as it was in the work of
Planck (1912) and Poincar\'{e} (1911, 1912), who traced the origin of the
new statistics to the discreteness of the energy. (More generally, I shall
trace it to the discreteness of the measure to be used, in equilibrium
conditions, on Hilbert space.)

\section{Against classical indistinguishability}

The clearest argument in the literature against classical
indistinguishability is that the principle is not needed (what I shall call
the `dispensability argument'); a thesis due to Ehrenfest and Trkal (1920),
and subsequently defended by van Kampen (1984). The objection that classical
indistinguishability is incoherent is more murky, and has rarely been
defended explicitly; for that I consider only van Kampen (1984) and Bach
(1997).

Ehrenfest and Trkal considered the equilibrium condition for molecules
subject to disassociation into a total $N^{\ast }$ of atoms, whose number is
conserved, with recombination into different possible numbers $N,N^{\prime
},N^{\prime \prime }...$ of molecules of various types. The upshot: a
contribution $k\log (N^{\ast }!/N!N^{\prime }!...)$ to the total entropy
which, when written as a sum of entropies for each type of molecule,
supplied in each case the necessary division by $N!,$ $N^{\prime }!$... etc.
(but with a remaining overall factor $N^{\ast }!$ - a number, however, that
did not change, so contributing only a constant to the entropy ). As a
result one has extensive molecular entropy functions (albeit a non-extensive
entropy function for the atoms), and can determine the equilibrium
concentrations of the molecules accordingly.

Van Kampen's argument was similar, and in certain respects simpler; let us
go into this in more detail. Consider a gas of $N^{\ast }$ particles,
defined as a canonical ensemble with the probability distribution:%
\begin{equation*}
W(N^{\ast },q,p)=f(N^{\ast })e^{-\beta H(q,p)}
\end{equation*}%
where $(q,p$) are coordinates on the $6N^{\ast }$ dimensional phase space
for a system of $N^{\ast }$ particles (with $f$ as yet undetermined). Let
the $N^{\ast }$ particles lie in volume $V^{\ast }$, and consider the
probability of finding $N$ with total energy $E$ in the sub-volume $V$ (so $%
N^{\prime }=N^{\ast }-N$ are in volume $V^{\prime }=V^{\ast }-V$). Assuming
the interaction energy between particles in $V^{\prime }$ and $V$ is small,
the Hamiltonian $H_{N^{\ast }}$ of the total system can be written as the
sum $H_{N}+H_{N^{\prime }}$ of the two subsystems. It then follows that the
probability density $W(N,q,p)$ for having $N$ particles at the point $%
(q,p)=q_{1},p_{1};q_{2},p_{2};....;q_{N},p_{N}$ should be calculated by the
procedure: first select $N$ out of the $N^{\ast }$ particles to be located
in $V$, and then integrate over all possible locations of the remaining $%
N^{\ast }-N$ particles in $V^{\prime }$, and repeat, allowing for different
selections. The result is:%
\begin{equation}
W(N,q,p)=const.\binom{N^{\ast }}{N}e^{-\beta
H_{N}(q,p)}\dint\limits_{V^{\prime }}e^{-\beta H_{N^{\prime }}(q^{\prime
},p^{\prime })}dq^{\prime }dp^{\prime }.
\end{equation}%
If one now goes to the limit $N^{\ast }\rightarrow \infty $, $V^{\prime
}\rightarrow \infty $, at constant density $N^{\ast }/V^{\prime }$ one
obtains the grand canonical distribution:%
\begin{equation*}
W(N,q,p)=const.\frac{z^{N}}{N!}e^{-\beta H_{N}(q,p)}
\end{equation*}%
where $z$ is a function of that density and of $\beta $, with the required
division by $N!$.

In this derivation the origin of the $1/N!$ is clear; it derives from the
binomial in Eq.(1) in the limit $N^{\ast }\rightarrow \infty $, i.e.:%
\begin{equation*}
\binom{N^{\ast }}{N}=\frac{N^{\ast }!}{(N^{\ast }-N)!N!}\rightarrow \frac{%
N^{\ast N}}{N!}
\end{equation*}%
where the integral over the volume $V^{\prime }$ supplies a further factor $%
V^{\prime N^{\ast }-N}$ (i.e. $const.V^{\prime -N}$). Thus\ it is only 
\textit{because} permutations of the $N^{\ast }$ particles yield physically
distinct states of affairs that one must divide through by $N!$ (to factor
out permutations that do not interchange particles inside $V$, with those in 
$V^{\prime }$). \ The same point applies to the model of Ehrenfest and
Trkal:\ it is only \textit{because} permutations of atoms are assumed to
lead to distinct states of affairs, that one must factor out permutations
that merely swap atoms among the same species of molecules.

Why so much work for so little reward? Why not simply assume that the
classical description be permutable (i.e. that points of phase space related
by a permutation of all $N^{\ast }$ particles represent the same physical
situation)? That, essentially, is what Gibbs' proposed. Van Kampen
considered the matter in the following terms:

\begin{quotation}
One could add, as an aside, that the energy surface can be partitioned in $%
N! $ equivalent parts, which differ from one another only by a permutation
of the molecules. The trajectory, however, does not recognize this
equivalence because it cannot jump from one point to an equivalent one.
There can be no good reason for identifying the $Z-$star [the region of
phase space picked out by given macroscopic conditions] with only one of
these equivalent parts. (van Kampen 1984, p.307).
\end{quotation}

\noindent (I shall come back to this argument somewhat later.) Gibbs' views
to the contrary he found `somewhat mystical' (van Kampen 1984, p.304).
Moreover:

\begin{quotation}
\noindent Gibbs argued that, since the observer cannot distinguish between
different molecules, "it seems in accordance with the spirit of the
statistical method" to count all microscopic states that differ only by a
permutation as a single one. Actually it is exactly opposite to the basic
idea of statistical mechanics, namely that the probability of a macrostate
is given by the measure of the Z-star, i.e. the number of corresponding,
macroscopically indistinguishable microstates. As mentioned...it is
impossible to justify the N! as long as one restricts oneself to a single
closed system. (van Kampen 1984, p.309).
\end{quotation}

\noindent These are the incoherence arguments, as we have them from van
Kampen.

The dispensability argument can be challenged head on. The extensivity of
the entropy, if it can be secured, even for contexts in which it has no
direct experimental meaning, hardly counts against the metaphysics, or
philosophical point of view, or physical interpretation, that underwrites
it; for it is in all cases desirable. Certainly it is possible to define a
classical thermodynamic entropy function that is extensive, and demarcates
precisely the thermodynamically allowed transformations of initial into
final states (whether or not by a quasi-static process), even for closed
systems, let the statistical mechanical account of it fall where it will.
Using the methods just outlined, one will be hard put to identify the
`ur-particles', whose number strictly does not change, in all physical
cases. The analysis of Ehrenfest and Trkal relied on the immutability of
atoms, but why not, as countenanced by Lieb and Yngvason (1999 p.27),
contemplate nuclear interactions as well? So long as initial and final
states are comprised of non-interacting systems, classically describable,
each in a (thermal) equilibrium state, classical thermodynamical principles
apply, however violent (and non-classical) the transformations that connect
them. These principles ensure the existence of entropy functions, additive
and extensive for each constituent classical equilibrium subsystem, now
matter how various. There is no difficulty in defining the latter, in
classical statistical mechanics, assuming permutability, but it is far from
clear how this may be achieved (or even that it should be possible at all)
when it is rejected, and one restricts oneself to the methods of Ehrenfest,
Trkal, and van Kampen. \footnote{%
Elements of this argument are due to Justin Pniower (see his (2006) for his
somewhat stronger claim that extensivity is indeed an empirically
falsifiable principle).}

Van Kampen's incoherence arguments were more rhetorical. It is true that
unobservability \textit{per se} is not a good reason, in statistical
mechanics, for identifying microscopic configurations; but Gibbs said only
that 'if the particles are \textit{regarded as indistinguishable}, it seems
in accordance with the statistical method...' (Gibbs 1902, p.187)), and for
a further indication of what he meant by the latter, his conclusion was that
`the question is one to be decided in accordance with the requirements of
practical convenience...' (p.188). Gibbs spoke as a pragmatist, not as a
positivist, nor as someone muddled on method.

I will approach van Kampen's remaining incoherence argument indirectly.
Alexander Bach, whose monograph \textit{Indistinguishable} \textit{Classical
Particles} goes a long way to rehabilitating the concept of
indistinguishability in classical statistical mechanics, voiced a related
objection that he himself found compelling. As a result, he thought it
important to distance his concept of classical indistinguishability from
this other, indefensible kind. The latter takes particle
indistinguishability all the way down to the microscopic details of
individual particle motions, whereas, according to Bach, it ought to concern
only statistical descriptions (probability measures). In the sense Bach
intended this restriction, it is simply not the same concept as
indistinguishability in quantum mechanics,\footnote{%
It is closer to De Finetti's concept of `exchangeability', called `purely
classical' by van Fraassen (1991, p.414).} \ which does go all the way down
to the microscopic level and the details (such as they are) of individual
particle motions. If Bach were right on this point, the concepts of
classical and quantum indistinguishability would differ fundamentally.

Bach is led to this view because:

\begin{quotation}
\noindent \textbf{Indistinguishable Classical Particles Have No Trajectories}%
. The unconventional role of indistinguishable classical particles is best
expressed by the fact that in a deterministic setting no indistinguishable
particles exist, or - equivalently - that indistinguishable classical
particles have no trajectories. Before I give a formal proof I argue as
follows. Suppose they have trajectories, then the particles can be
identified by them and are, therefore, not indistinguishable. (Bach 1997
p.7).
\end{quotation}

Bach's formal proof proceeds by identifying the coordinates of such a pair
(in 1-dimension) as an extremal of the set of probability measures $%
M_{+}^{1}(\mathbf{R}^{2})$, from which the `diagonal' $D=\{<x,x>\in \mathbf{R%
}^{2},x\in \mathbf{R}\}$ is deleted (because the particles are assumed
impenetrable); and by characterizing classical indistinguishability as a
feature of the state, \ namely states in $M_{+,sym}^{1}(\mathbf{R}^{2})$,
whose extremals are of the form 
\begin{equation*}
\mu _{x,y}=\frac{1}{2}\left( \delta _{<x,y>}+\delta _{<y,x>}\right) ,\text{ }%
<x,y>\in \mathbf{R}^{2}\backslash D
\end{equation*}%
(i.e. states concentrated on the points $<x,y>,$ $<y,x>$, $x\neq y$). It
concludes with the observation that no such symmetric state is an extremal
of $M_{+}^{1}(\mathbf{R}^{2})$, hence no such state assigns coordinates to
the particle pair.

But why not say instead that the coordinates of classical indistinguishable
particles on the contrary attach to points in the \textit{reduced state space%
}? I.e., for two particles in $1-$dimension, they are extremals not of $%
M_{+}^{1}(\mathbf{R}^{2})$, but of $M_{+}^{1}(\mathbf{R}^{2}/\Pi _{2})$,
where $\mathbf{R}^{2}/\Pi _{2}$ is the space obtained by identifying points
in $\mathbf{R}^{2}$ that differ only by a permutation. Passing to this
quotient space defeats Bach's formal argument. This can also provide a
starting point for the definition of the quantum theory of indistinguishable
particles (by quantization on the reduced configuration space).\footnote{%
As shown by Leinaas and Myrheim (1977). It is of interest that of the two
objections to Schr\"{o}dinger's use of functions on configuration space made
by Einstein at the fifth Solvay conference, one of them was that points
related by permutations were not identified (Einstein (1928); whether he
would have welcomed Leinaas and Myrheim's clarification is not so clear).}

We should locate clearly our point of difference with Bach. His argument
that identical particles cannot have trajectories, for otherwise particles
would be identifiable by them, was intended to show not that classical
indistinguishability makes no sense, but that it only makes sense if
description- relative (and, indeed, is restricted to a level of description
that does not describe individual trajectories). Hence his equation:

\begin{quotation}
\noindent Indistinguishability = Identity of the Particles + Symmetry of the
State (Bach 1997 p.8).
\end{quotation}

\noindent We can agree with Bach that indistinguishability is a matter of
the symmetry (permutability) of the description, but not with his further
point, that a symmetric description is impossible if it is so detailed as to
specify the trajectories. For why not allow that an equivalence class of
trajectories in configuration space (under the equivalence relation `is a
permutation of') indeed specify a single trajectory? \ - not, of course, in
configuration space, but in \textit{reduced} configuration space. We are
clearly going round in circles.

The case against classical indistinguishability that it is unnecessary is
moot, that it is incoherent is question-begging. What of van Kampen's
criticism that it is unmotivated? Granted that macroscopic unobservability
is not in general a condition, in statistical mechanics, for identifying
putatively distinct states, there remains another condition which is - which
is in fact much more universal. Indeed, it can be formulated and applied
across the gamut of physical theories. The condition is this: insofar as
permutations are mathematical symmetries of the equations, adequate to a
given set of applications (for a closed system), then they should be treated
just like any other group of symmetries - that is, points in the state space
for such systems related by the symmetry transformation should be identified
(and we should pass to the quotient space). This point is a familiar one in
the context of space-time symmetries, for example translations in space,
where the quotient space is the space of relative distances. Why not treat
permutations just like any other symmetry group, and factor them out
accordingly?\footnote{%
For the general method and its rational, see my (2003a,b).}

\section{Demystifying Classical indistinguishability}

The answer, presumably, is that we surely \textit{can} single out classical
particles uniquely, by reference to their trajectories. But there is a key
objection to this line of thinking:\textit{\ so can quantum particles, at
least in certain circumstances, be distinguished by their states}. No matter
whether the state is localized or not, the `up' state of spin, for example,
is distinguished from the `down', and may well be distinguished in this way
over time. In such cases, particle properties can be used as names.\footnote{%
This point was earlier made by Shankar (1980 p.283-88, particularly p.284;
I\ am grateful to Antony Valentini for bringing this to my attention). I
will come back to this matter in the final section.}

In the case of fermions it might even be thought that such an identification
is always possible. Thus Pauli recounts his discovery:

\begin{quotation}
On the basis of my earlier results on the classification of spectral terms
in a strong magnetic field the general formulation of the exclusion
principle became clear to me. The fundamental idea can be stated in the
following way: The complicated numbers of electrons in closed subgroups are
reduced to the simple number one if the division of the groups by giving the
values of the four quantum numbers of an electron is carried so far that
every degeneracy is removed. An entirely non-degenerate energy \ level is
already `closed', if it is occupied by a single electron: states in
contradiction with this postulate have to be excluded. (Pauli 1946 p.29).
\end{quotation}

\noindent Electrons may be simply identified by their quantum numbers (and
as such the permutations have nothing to act upon). As Stachel (2002) has
recently remarked, extending Einstein's famous `hole argument' to a purely
set theoretic setting (whereupon the symmetries analogous to diffeomorphisms
are permutations), one can talk of the pattern positions themselves as the
objects (which are not themselves permutable, no more than sets of quantum
numbers); he too recommends that we identify electrons by their quantum
numbers.

This amounts to identifying electrons as 1-particle states. Of course there
are plenty of situations where (because energy degeneracies are \textit{not}
always eliminated) this does not suffice to point to any unique electron,%
\footnote{%
A point recently made by Pooley (2006); here Pooley also argues against the
similarity of permutation symmetry in quantum mechanics with general
covariance in space-time theories, a point I shall come back to at the end.}
but these concern further symmetries, unrelated to permutations \textit{per
se}; symmetries which may also be present in the classical case and lead to
exactly the same difficulty (in such situations one cannot refer to a unique
classical trajectory either).\footnote{%
But in every case one can still \textit{discern} between the electrons, or
classical particles; see my (2003a, 2003b, and particularly 2006) for
further discussion.} The real distinction in the two cases is this: in
quantum mechanics, given an (anti)symmetrized state constructed from a given
set of orthogonal vectors $\{\varphi _{k}\},$ $k=1,..,N$, one can
individuate one particle from the remaining $N-1$ by its state, and one can
in principle, when the Hamiltonian factorizes, track the time evolution of
that particle (that state); but nothing comparable is possible if the state
is a \textit{superposition} of such (anti)symmetrized states.

That marks a profound difference from the classical case, but it does not
affect the comparison we are concerned with: a state of definite occupation
numbers is nevertheless permutation invariant, and the particles it
describes are still indistinguishable. To this the classical analog is
clear: just as we may speak of quantum states, rather than particles having
states, so we may speak of classical trajectories, rather than particles
having trajectories. But equally, if we do talk of the particles (that may
be in various states, or have various trajectories), that are otherwise the
same, then we should do so identifying permutations of particles among
states, for there is no further fact as to which particle is in which state,
or which has which trajectory, relevant to the dynamics. Returning to Van
Kampen's `incoherence' argument that `the trajectory... does not recognize
this equivalence because it cannot jump from one point to an equivalent one'
draws its effect, so far as it goes, from mixing the two kinds of
description. Our conclusion, again: indistinguishability (permutability,
invariance under permutations) makes just as much sense classically as it
does in quantum mechanics.

The matter can even be pushed as a point of logic, for the requirement of
indistinguishability, understood as permutability, would seem to make no
difference to a language that is void of proper names. Help yourself to such
a language, bracketing for the time being any scruples you may have on its
ultimate adequacy; then you are in much the same position as if you had
restricted yourself to complex predicates totally symmetric in all of their
arguments (at least if you restrict attention to finite numbers of objects).
For it can be proved (see the Appendix):\footnote{%
For further discussion, see Saunders (2006).}

\begin{quotation}
\noindent \textbf{Theorem}: Let $L$\ be a first-order language without any
proper names ($0-$ary function symbols). Let $T$ be any $L-$sentence
satisfiable only in models of cardinality $N.$ Then there is a totally
symmetric predicate $Gx_{1}...x_{N}$ $\in L$ such that $\exists
x_{1}...\exists x_{N}Gx_{1}...x_{N}$ is logically equivalent to $T_{.}$
\end{quotation}

\noindent The intuitive point is indeed very simply made if we consider only
purely existential sentences, like $\exists x_{1}...\exists
x_{N}Fx_{1}...x_{N}$, which is obviously logically equivalent to $\exists
x_{1}...\exists x_{N}\dbigvee_{k=1}^{N}Fx_{1}...Fx_{N}$, whatever the
predicate $F$. And sentences like this are, one would have thought,
perfectly sufficient to describe the configuration of a system of particles
in space.

But that makes the restriction (if it amounts to no more than the renouncing
of names) seem \textit{purely} metaphysical (and note that it applies
equally to descriptions of \textit{non-identical} particles).\footnote{%
If `essential' attributes of particles - charge and mass and so on - were
also specified by the state, one would indeed have a theory in which all
particles whatsoever are permutable. \ (I shall come back to this point in
Section 5.)} Indeed, according to Huggett (1999), the principle of
indistinguishability is none other than \textit{antihaecceitism}, an old
doctrine of scholastic philosophy, and one that is surely devoid of
empirical significance. His conclusion was endorsed by Albert (2000):

\begin{quotation}
There is a certain fairly trivial sense in which it ought to have been
obvious from the outset (if we had stopped to think about it) that the facts
of thermodynamics cannot possibly shed any light on the truth or falsehood
of the doctrine of Haecceisstism. The question of the truth or falsehood of
the second law of thermodynamics is (after all) a straightforwardly
empirical one; and the question of Haecceisstism, the question (that is) of
whether or not certain observationally identical situations are identical
simpliciter, manifestly is not. Nevertheless, it might have turned out that
the statistical-mechanical account of thermodynamics is somehow radically
simpler or more natural or more compelling or more of an explanatory success
when expressed in a Haecceisstic language than it is when expressed in a
non-Haecceisstic one. And the thing we've just learned (which seems to me
substantive and non-trivial and impossible to have anticipated without doing
the work) is that that is not the case. ( p.47-48)
\end{quotation}

\noindent But whilst I have some sympathy with Huggett's equation, it should
be obvious from the discussion of Section 2 that all is not well with this
way of putting it. To suggest, as Huggett did (citing van Kampen (1984)),
that extensivity of the entropy is only a `convention', is clearly
unsatisfactory. But that to one side, it is obvious that permutability 
\textit{can} have empirical question, indeed \textit{straightforward }%
empirical consequences, for if the state-space is Hilbert space, rather than
phase space, it forces a change in particle statistics! How can a change in
metaphysics have that consequence? And with that we are back to our puzzle:\
What is responsible for the difference between quantum and classical
statistics?

Permutability, we should conclude, is not a metaphysical principle, nor
merely a convention; if not in itself an empirical claim, it makes a
contribution to others that are. But we need not pursue the question of the
precise status of this principle, given only that it makes classical sense;
whereupon we are returned to the puzzle as stated.

\section{The explanation for quantum statistics}

To begin at the beginning:

\begin{quotation}
The distribution of energy over each type of resonator must now be
considered, first, the distribution of the energy $E$ over the $N$
resonators with frequency $\upsilon $. If $E$ \ is regarded as infinitely
divisible, an infinite number of different distributions is possible. We,
however, consider - and this is the essential point - $E$ to be composed of
a determinate number of equal finite parts and employ in their determination
the natural constant $h$= 6.55$\times 10^{-27}$ erg sec. This constant,
multiplied by the frequency, $\upsilon $, of the resonator yields the energy
element \ $\Delta \epsilon $ in ergs, and dividing $E$ by $h\upsilon $, we
obtain the number $P$, of energy elements to be distributed over the $N$
resonators. (Planck 1900 p.239).
\end{quotation}

It is noteworthy that permutability (indistinguishability) seemed perfectly
natural to Planck in this setting: for why distinguish situations in which
one entity is allocated to one resonator, rather than another, when the
entity is merely an `energy element'? Boltzmann likewise identified
permutations of energy elements, both in his 1868 derivation and that of
1877 (using the combinatorics factor below) - but differed in the crucial
respect that he took the limit in which the energy elements went to zero.%
\footnote{%
I refer to Bach (1990) for a detailed study of this history.}

How many ways can $P$ energy elements be arranged among $N$ resonators? This
question is important, if each such arrangement is equiprobable (as we
assume). Call the number $W_{I}$. For it Planck took from Boltzmann the
expression:%
\begin{equation}
W_{I}=\frac{(P+N-1)!}{P!(N-1)!}.
\end{equation}%
For (this derivation is due to Ehrenfest) an arrangement can be given as a
sequence of symbols (where $n_{i}\in \{0,1,2,...,P\}$, $i=1,..,N$): 
\begin{equation*}
\underset{n_{1}}{\underbrace{p...p}}|\underset{n_{2}}{\underbrace{p....p}}%
|.....|\underset{n_{N}}{\underbrace{p...p}}
\end{equation*}%
of which there are $P$ \ symbols `$p$' in all (so $\sum_{s=1}^{N}n_{s}=P$),
and $N-1$ symbols `$|$'. Given such a sequence one can say exactly how many
energy elements $n_{k}$ are in the $k$-th cell (the `occupation number' of
each cell), but not which energy element is in which cell. If no such
further facts are either relevant or available, $W_{I}$ is then simply the
number of distinct sets of occupation numbers (of distinct arrangements, in
Planck's sense). There are $(P+N-1)!$ permutations of $P+N-1$ symbols in
all, but of these, those which only shuffle `$p$' s among themselves, or `$|$%
's among themselves, do not give us a new set of occupation numbers; so we
must divide by $P!$ and by $(N-1)!$.

Now adopt a change in notation and physical picture; let the $P$ quanta be
called `particles', and the $N$ oscillators `cells in phase space', with the
new notation `$N$' and `$C$' respectively. The question may now seem to
arise as to which particle is in which cell; to which the answer is there
are $C^{N}$ possible choices, one for each set of `occupation numbers' $%
n_{k} $, $k=1,..,C$, with a multiplicity to allow for permutations of
particles among different cells. In this time-honoured way obtain:\footnote{%
The second equality is obvious by inspection. It is a special case of a more
general theorem (the multinomial theorem), which says that for arbitrary
quantities $z_{1}...z_{C}$ 
\begin{equation*}
(z_{1}+..+z_{C})^{N}=\sum_{\substack{ \text{all }C-\text{tuples of integers }%
n_{1}...n_{C}  \\ \text{s.t. }\sum_{s=1}^{C}n_{s}=N}}\frac{%
N!z_{1}^{n_{1}}...z_{C}^{n_{C}}}{n_{1}!...n_{C}!}
\end{equation*}%
\par
\noindent As defined below, Botzmann's count of complexions is obtained for $%
z_{1}=...=z_{C}=1$, his volume measure for $z_{1}=..=z_{C}=\tau $. (For this
and the combinatoric expressions that follow, see e.g. Rapp (1972).)}

\begin{equation}
W_{D}=C^{N}=\dsum\limits_{\substack{ \text{occupation numbers}  \\ \text{%
s.t. }\sum_{s=1}^{C}n_{s}=N}}\frac{N!}{n_{1}!...n_{C}!}.
\end{equation}

There is one further complication: the numbers $C,$ $N$ are associated with
particular regions of phase space, parameterized by other variables (usually
the energy). Thus, in the case of the Planck distribution, by the frequency
(so $C_{k}$ cells in the $k^{th}$ frequency range, etc.). If $E_{k}$ is the
energy of the $k^{th}$ region, then $C_{k}$ is the corresponding degeneracy
(number of cells all in that energy range). Let $N_{k}$ particles lie in
region $k$, and let there be $j$ regions in all; then the constraint on the
total energy $E_{tot}$ is that $\sum_{k=1}^{j}N_{k}E_{k}=E_{tot}$; if
particle number too is conserved (with $N_{tot}$ particles in all), then the
total number of arrangements is:\ 
\begin{equation}
W_{D}=\dsum\limits_{\substack{ \text{sequences }N_{1}...N_{j}  \\ \text{s.t. 
}\sum_{s=1}^{j}N_{s}=N_{tot}\text{, }\sum_{s=1}^{j}N_{s}E_{s}=E_{tot}}}\frac{%
N_{tot}!}{N_{1}!...N_{j}!}\dprod\limits_{k=1}^{j}C_{k}^{N_{k}}.
\end{equation}%
Here the permutation factors arise as distinct regions of phase space
correspond to different choices as to which $N_{1}$ (of $N_{tot}$ particles)
are assigned to region $1$ (with $C_{1}$ cells), which $N_{2}$ (of $%
N_{tot}-N_{1}$) to $2$ (with $C_{2}$ cells), and so on.

Eq(4) was also written down by Boltzmann, in his 1877 memoir. Contrast it
with the analogous expression for $W_{I}$:%
\begin{equation}
W_{I}=\dsum\limits_{\substack{ \text{sequences }N_{1}...N_{j}  \\ \text{s.t. 
}\sum_{s=1}^{j}N_{s}=N_{tot},\text{ }\sum_{s=1}^{j}N_{s}E_{s}=E_{tot}}}%
\dprod\limits_{k=1}^{j}\frac{(N_{k}+C_{k}-1)!}{N_{k}!(C_{k}-1)!}.
\end{equation}%
It is a spurious simplification of (4) to suppose that the degeneracy of
each energy $E_{k}$ is unity (i.e. $C_{k}=1$ for each $k)$, and to go on to
identify the $N_{k}$'s with the occupation numbers $n_{k}$ of (3). Under
that assumption, the limiting agreement between (4) and (5) is rather hard
to see. It is that in the limit in which $C_{k}\gg N_{k}$%
\begin{equation}
\frac{(N_{k}+C_{k}-1)!}{N_{k}!(C_{k}-1)!}\approx \frac{C_{k}^{N_{k}}}{N_{k}!}
\end{equation}%
whereupon $W_{I}$ $\approx W_{D}/N_{tot}!$. But away from this limit, the
two are quite different; moreover, the quantity $W_{D}/N_{tot}!$ is not a
combinatorial count of anything (it is not an integer). Rather, we should
interpret it as the expression:%
\begin{equation}
\frac{W_{D}}{N_{tot}!}=\frac{1}{\tau ^{N_{tot}}}\dsum\limits_{\substack{ 
\text{sequences }N_{1}...N_{j}  \\ \text{s.t. }\sum_{s=1}^{j}N_{s}=N_{tot}%
\text{, }\sum_{s=1}^{j}N_{s}E_{s}=E_{tot}}}\dprod\limits_{k=1}^{j}\frac{%
(C_{k}\tau )^{N_{k}}}{N_{k}!}
\end{equation}%
\textit{\ }i.e. \textit{the reduced phase space volume }(in the units $\tau
^{N_{tot}}$) corresponding to the stated constraints on $N$ and $E$. Each
term in the product is the reduced $N_{k}-$particle phase space volume
corresponding to the $C_{k}$ cells in the 1-particle phase space, each of
volume $\tau .$

The breakdown of the approximation (6) is responsible for the entire
difference between classical and quantum statistical equilibria. Evidently
to understand it we need only investigate it for a single (arbitrary) phase
space region $k$ \noindent (so from this point on we drop the subscripts on `%
$N_{k}$' and `$C_{k}$'.) For low numbers the approximation can be
illustrated graphically. We take the simplest case of $N=2$ particles in $1-$%
dimension (so with a $4-$dimensional phase space). First the distinguishable
case.

\paragraph{Distinguishable particles}

Divide each 1-particle phase space into $C$ cells, as did Boltzmann, say $C=3
$. Then there are $C^{N}=3^{2}=9$ different ways the two particles can be
distributed in this region of the $2-$particle phase space. Thus, supposing
the two particles are named `$a$', `$b$', the region in $\Gamma ^{2}$
corresponding to the arrangement in which particle $a$ is in cell \ 2 and
particle $b$ is in cell 3 is the region shaded (Fig.1, suppressing one
dimension of $\Gamma ^{1}$). If each such arrangement is equiprobable, one
obtains Maxwell-Boltzmann statistics.\bigskip 


\begin{center}
Fig.1:\ Distinguishable particles
\end{center}

On the alternative way of putting it, in terms of phase space volume, if $%
\tau $ is the size of each cell in the 1-particle phase space, the volume of
the $N-$particle phase space is $(C\tau )^{N}$, or $\tau ^{N}$ times the
number of all arrangements of $N$ particles in $\Gamma ^{N}$. That is, we
may equally take Boltzmann's thermodynamic probability as phase space volume
(Boltzmann himself often spoke of it in this way).

These two quantities, the count of arrangements, and their corresponding
volume, are quite generally proportional to the count of available states in
the Hilbert space of $N$ distinguishable particles, corresponding to the
classical phase space region $\Gamma ^{N}$, each with a 1-particle Hilbert
space of $C$ dimensions. For $N=2$ and $C=3$, as above, we have a 2-particle
Hilbert space $H^{2}=H^{1}\otimes H^{1}$, where each 1-particle space is
spanned by three orthogonal states $\varphi _{1}$, $\varphi _{2},\varphi
_{3} $. There are again $C^{N}=3^{2}=9$ orthogonal two-particle state
spanning $H^{2}.$

There is therefore one orthogonal 2-particle state in $H^{2}$ (represented
by dots in Fig 1) for each arrangement in $\Gamma ^{2}$, each with the same
phase-space volume by Boltzmann's assumption, and each with the same Hilbert
space measure (counting each state - each dot - as equiprobable). Under
these assumptions, the measures on the quantum and classical state spaces
are the same (the number of dots is proportional to the total phase space
volume); so distinguishable quantum particles also obey Maxwell-Boltzmann
statistics. Of course there are other important differences, notably, that
the entropy no longer has an arbitrary additive constant (corresponding to
the arbitrary choice of the unit $\tau $); its value at absolute zero, in
particular, is $Nk\ln C_{0}$, where $C_{0}$ is the dimensionality of the
subspace of lowest energy $E_{0}$. For another, that for sufficiently small
temperatures, only particles in the lowest energy states contribute to the
specific heats of solids (as discovered by Einstein in 1907) - but this
bares more on the discreteness of the spectrum of the energy.

\paragraph{Indistinguishable particles}

In the case of classical indistinguishable particles we should use instead
the measure on the reduced phase space for this region, i.e. $\Gamma
^{N}/\Pi _{N}$. For $N=2,$ $C=3$ this amounts to going over to Fig.2. The
volume goes down from $(C\tau )^{N}$ to $(C\tau )^{N}/N!$; the number of
arrangements \textit{also} goes down, \textit{but not by the same factor}.
And correspondingly, the volume of each arrangement is no longer the same.

\begin{center}
Fig.2\bigskip : Indistinguishable particles
\end{center}

This point is obvious by inspection of Fig.2: the volume is now $(C\tau
)^{N}/N!=9\tau ^{2}/2$, but of course there are not $C^{N}/N!$ $=4.5$
arrangements - rather, there are precisely six different ways of
distributing two indistinguishable particles over three $1$-particle cells,
without regard for which is in which cell (one for each dot in Fig.2); but
clearly the reduced phase space volumes of three of the arrangements are
twice those of the others (the ones along the diagonal - this is why in the
classical case the statistics remains Maxwell-Boltzmann. Only the ratios of
volumes, the relative probabilities, matter to the statistics). Suppose $C=2$
(so ignoring the top row and rightmost column); take in illustration two
fair coins, with the two regions of the $1-$coin phase space labelled `$H$'
and `$T$' respectively; then the outcome $\{H,T\}$ (an unordered pair) is
twice as likely as either $\{H,H\}$ or $\{T,T\}$, just as for
distinguishable coins.

The count of distinct arrangements is given by (2). In quantum mechanics,
this count goes over (for any basis) to the count of orthogonal totally
symmetrized states in Hilbert space (the dimension of the subspace
corresponding to the macroscopic constraints). In our example, this includes
the $3-$dimensional subspace spanned by the vectors $\varphi _{k}\otimes
\varphi _{k},$ $k=1,2,3$, as before, but now directly summed not with the $6$%
-dimensional subspace spanned by $\varphi _{k}\otimes \varphi _{j}$, $j\neq
k $, but with the $3$-dimensional subspace spanned by $\varphi _{k}\otimes
\varphi _{j}+\varphi _{j}\otimes \varphi _{k}$, $j\neq k$. The crucial
difference with the classical case is: \textit{there is no other measure on
the state space but this}. And using this measure, the diagonals must have
the \textit{same} probability as the off-diagonals - therein lies the
difference with classical theory (and the reason why, for two quantum coins,
the probabilities for $\{H,T\}$, $\{H,H\}$. and $\{T,T\}$ are all the same).
Arriving at a quantity like $(C\tau )^{N}/N!$, rather than $%
(N+C-1)!/N!(C-1)! $, is not an option.\footnote{%
My thanks to David Wallace for conversations on this point.}

It is worth making this point again in terms of the occupation numbers. For
each arrangement of distinguishable particles, there are $N!/n!...n_{C}!$
distinct assignments of the $N$ particles over $C$ cells, so as to place $%
n_{k}$ in cell $k$, $k=1,...,C$. The total number of such arrangements is $%
C^{N}$, which as we have seen (Eq.(3)) is:

\begin{equation*}
\dsum\limits_{\substack{ \text{occupation numbers}  \\ \text{s.t. }%
\sum_{k=1}^{C}n_{k}=N}}\frac{N!}{n_{1}!...n_{C}!}=C^{N}.
\end{equation*}%
Why then, if division by $N!$ compensates for unwanted permutations, do we
not obtain in this way the same result as did Planck? In fact, to get the
Planck expression, the factor $N!/n_{1}!...n_{C}!$ is set equal to \textit{%
one}, weighting each Planck arrangement the same, thus obtaining:%
\begin{equation*}
\dsum\limits_{\substack{ \text{occupation numbers}  \\ \text{s.t. }%
\sum_{k=1}^{C}n_{k}=N}}1=\frac{(N+C-1)!}{N!(C-1)!}.
\end{equation*}%
That you should not do, thinking classically, if the volume is what matters,
and you are going to identify phase space points related by permutations.
For in that case the volumes of arrangements along the diagonals of the
reduced phase space (with occupation numbers greater than one) should be
weighted\textit{\ less} than all the rest. Since the relative weights are
all that matter to the statistics, using the factor $N!/n_{1}!...n_{C}!$, or 
$1/n_{1}!...n_{C}!$ (dividing by an overall factor of $N!$, as one should),
makes no difference. In comparison to this, quantum mechanics, \textit{%
weighting them equally}, increases their weights in comparison to their
classical values. This explains the comment, often made, that particles
obeying Bose-Einstein statistics have the tendency `to condense into groups'
(Pauli 1973 p.99).

Quantum mechanically there is no volume measure, and no reason to weight one
set of occupation numbers differently from any other. Classically, one might
think \textit{both} options are on the table: the (integral) count of
arrangements, given by Planck's expression, of the (non-integral) volume of
reduced phase space, as given by the corrected Boltzmann expression. But the
former, unlike the latter, depends crucially on the size of the elementary
volume $\tau ;$ if there is to be a departure from classical statistics on
this basis, it will require the existence of a fundamental unit of phase
space volume, with the dimensions of action.\footnote{%
A consideration that applies to those, like Costantini (1987), who have
claimed to explain quantun statistics in classical terms.} The very
discrepancy between Planck's expression and the reduced phase space volume
(the quantity $C^{N}/N!)$ disappears, as it must, as $\tau \rightarrow 0$,
as inspection of Fig.2 makes clear (imagine the triangle partitioned into
much \textit{smaller} squares; then the number of states - the number of
dots - \textit{is} approximately proportional to the area). Planck's
combinatorial count, multiplied by $\tau ^{N}$, is an increasingly good
approximation to the volume as $\tau $ (or equivalently $1/C^{N}$) become
small; that is just the condition $C$ $\gg N$ considered previously, under
which quantum and classical statistics agree (where on average no particle
has the same 1-particle energy as any other, and the arrangements along the
diagonal are on average unoccupied).

For the same reason Fermi-Dirac statistics are undefinable in classical
terms, for the condition that no two particles are in the same 1-particle
state, or in the same cell of the 1-particle phase space, is only physically
meaningful if these cells have a definite size.\footnote{%
Note added in proof: a condition that can, admittedly, be formulated
independent of quantum mechanics (e.g. in terms of a lattice theory); as
explained by Gottesman (2005). }

\section{Addenda}

What explains the difference between classical and quantum statistics? The
structure of their state spaces: in the quantum case the measure is
discrete, the sum over states, but in the classical case it is continuous.%
\footnote{%
I do not suppose quantum interference phenomena more generally, traceable to
(anti)symmetrization of the state, are similarly explained (I\ am grateful
to Lee Smolin and Rafael Sorkin for pressing this point upon me).} This
makes a difference when one passes to the quotient space under permutations,
as we should for particles intrinsically alike.

Our purely logical theorem, that shows all objects (of any finite
collective) to be permutable (in a language without names) is evidently much
broader in scope, for there is no restriction to objects intrinsically the
same. There is also another long-standing tradition (due to Feynman 1965),
that explains the distinction between classical and quantum statistics in
terms of the possibility (or lack of it) of reidentification of particles
over time. Both raise questions over the adequacy of the explanation so far
given.

In fact the two are connected. For let us suppose that distinct particles
may be labelled by distinct properties that are constant in time. Take again
the example of two coins ($N=C=2),$ but suppose now that one coin is red ($r$%
) and the other green $(g)$. Taking `red', `green' as proper names, one has
distinguishable coins that may take on one of two phases ($H$ or $T$), and
correspondingly one has the unreduced phase space, similar to Fig.1, but now
a $2$ by $2$ grid (with `$r$' and `$g$' in place of `$a$' and `$b$'). But
assimilating colours to the phases instead, we pass to the case $C=4$ (the
four phases $\{H,r\}$, $\{H,g\}$, $\{T,r\}$ and $\{T,g\}$), and the coins
are again indistinguishable (with the reduced phase space of Fig.3).
However, supposing the colours are stable under each toss of the coins,
certain cells of the $2$ particle phase space are inaccessible (those in
which the two coins have the same colour, the regions shaded); the \textit{%
effective} phase space for the $2$ indistinguishable coins, consisting of
the unshaded boxes in Fig.3, gives back the original unreduced phase space
(where the colours function as names).

\begin{center}
Fig.3\medskip : Recovering distinguishable particles
\end{center}

This explains why differences in intrinsic particle properties (such as
mass, spin and charge), stable in time, are grounds for treating them as
distinguishable (with no need to (anti)symmetrize in Hilbert space). The
point about identification over time also falls into place; whatever the
criterion for each particle, it is \textit{ex hypothesis} stable in time and
shared with no other. In the classical case, where there are definite
trajectories, one can construct such properties by reference to a location
in phase space at a given time (so that a particle has that property if and
only if its trajectory passes through that location at that time). Quantum
mechanically, for a state of non-interacting particles of definite
occupation numbers at a given time (all of them $0$'s and $1$'s), the
procedure is the same (the orbits of $1$-particle states replace
trajectories); or, alternatively, in the solid state, taking particles as
lattice-sites, identifying particles by their position over time. In any of
these ways one can pass to a phase space or Hilbert space description in
terms of distinguishable particles, subject to Maxwell-Boltzmann statistics,
and a non-extensive entropy. The utility of any such description, however,
will depend on the ingenuity of the experimenter, to define operational
conditions for the reidentification of such particles over time. In the
solid state such conditions are plain; identification in terms of states is
also possible in the high-temperature limit (occupation numbers all $0$'s
and $1$'s), where talk of $1-$particle states amounts to talk of modes of
the quantum field (with excitation numbers all $0$'s or $1$'s) - this goes
some way to explaining Schr\"{o}dinger's \ result, that one can treat
Bose-Einstein particles in terms of waves obeying Maxwell-Botlzmann
statistics.\footnote{%
And suggests a corresponding account of fermions.} For the classical example
where one reidentifies particles over time by their trajectories, one needs
more fanciful conditions, say a Maxwell's demon able to keep track of the
individual molecules of a gas over time; that makes clear why one ought to
have an entropy of mixing, and hence a non-extensive entropy function, in
such circumstances.

A final comment. It may be objected that the treatment of
indistinguishability is different in quantum mechanics than classically, and
different from other classical symmetries like general covariance, precisely
because one does not, in quantum mechanics, and parastatistics to one side,
take an equivalence class of states as representing the physical situation;
one takes instead the symmetrized state, itself a vector in the unreduced
state space. Nothing comparable is available classically.

It is true that classical and quantum mechanics differ in this respect:
classically only very special states in the reduced state space are also to
be found in the unreduced space (and none at all if, following Bach, the
diagonals are omitted). But the more general point, that classically one
works not with a single invariant state, but with an equivalence class of
states in the unreduced state space, I take to be a reflection of something
still more fundamental: it is that whilst in both cases one can pass to the
quotient space, only in quantum mechanics is the topology preserved
unchanged (the space of symmetrized vectors is topologically closed, so it
itself a subspace). The topology of the classical quotient space, under
permutations, is in contrast enormously more complex than that of the
unreduced space (and, omitting the diagonals, is not even topologically
closed). Easier, then, classically, to work in the unreduced state space,
taking the equivalence class of points as representative of the physical
situation.\footnote{%
As observed by Gibbs: `For the analytical description of a specific phase is
more simple than that of a generic phase. And it is a more simple matter to
make a multiple integral extend over all possible specific phases than to
make one extend without repetition over all possible generic phases.' (Gibbs
1902 p.188).}

\bigskip \bigskip \bigskip \bigskip 

\noindent \bigskip {\Large Appendix\nolinebreak \nolinebreak \bigskip }

\begin{proof}
\nolinebreak Let $L^{+}$ differ from $L$\ only in the addition of countably
many names $a_{1},a_{2},..$. The proof proceeds as follows: we construct a
sequence of sentences $A$, $T_{1}$, $T_{2},$ $T_{S},$ where the last is of
the desired form, where $T_{1},T_{2}\in L^{+},$ and $A$ is the $L^{+}-$%
sentence $\dbigwedge\limits_{i,j=1,i\neq j}^{N}a_{i}\neq a_{j}\wedge \forall
x\dbigvee_{k=1}^{N}x=a_{k},$ satisfying:

i) $T\wedge A\vDash T_{1}$

(ii) $T_{S}\wedge A\vDash T_{2}$

(iii) $\vDash T_{2}\leftrightarrow T_{1}$

(iv) $T_{1}\vDash T$

(v ) $T_{2}\vDash T_{S}.$

\noindent Since any sentence $S$ in any first-order language has the same
truth value in models that differ only in their interpretations of
non-logical symbols not contained in $S$, the truth of $T,$ $T_{S}$ (which
contain no names)\ in a model of $L^{+}$ with universe $V$ is independent of
the assignment of names in $L^{+}$ to elements of $V$. It then follows from
(i) through (v) that $\vDash T\leftrightarrow T_{S}.$ For suppose $T_{S}$ is
true in $V$; if $A$ is also true, then by (ii),(iii),(iv), $T$ is true in $V.
$ Suppose $T_{S}$ is true and $A$ is false in $V$; choose a model $W$
identical to $V$ save in the interpretation of symbols not in $T_{S}$, $T$,
in which $A$ is true (it will be clear from the construction of $T_{S}$ that
it only has models of cardinality $N$, so such a model can always be found).
Then as before, $T$ is true in $W$; hence also in $V$. Thus $T_{S}\vDash T$.
The proof that $T\vDash T_{S}$ uses (i), (iii), (v), but is otherwise the
same.

It remains to prove (i) through (v). Define $T_{1}$ as $T\wedge A$ (so (i),
(iv) follow immediately). Without loss of generality, let $T$ be given in
prenex normal form, i.e. as a formula $Q_{n}...Q_{1}Fx_{1}...x_{n}$, $n\geq
1 $, where each $Q_{i}$ is either $\forall x_{i}$ or $\exists x_{i}$. Define
a sequence of sentences $T^{(1)},...,T^{(n)}$ by:

$T^{(k)}$ $\underset{def}{=}$ $%
Q_{1}...Q_{n-k}G^{(k)}x_{1}...x_{n-k}a_{1}...a_{N},$ for $1\leq k<n.$

$T^{(n)}\underset{def}{=}$ $G^{(n)}a_{1}...a_{N}.$

The predicates $G^{(1)},...,G^{(n)}$ are defined as follows. Let $\mathbf{[}k%
\mathbf{]}$ be $\dbigwedge $ if $Q_{k}$ is $\forall x_{k}$, and otherwise $%
\dbigvee $; for any predicate $Px_{1}...x_{j}...$, let $Px_{1}...\underset{j}%
{a_{k}}...$ denote the result of replacing every occurrence of $x_{j}$ in $P$
by $a_{k}.$ Then:\medskip

\noindent $G^{(1)}x_{1}...x_{n-1}a_{1}...a_{N}\underset{def}{=}\mathbf{[}n%
\mathbf{]}_{i-1}^{N}Fx_{1}...x_{n-1}\underset{n}{a_{i}}$

\noindent $G^{(k+1)}x_{1}...x_{n-(k+1)}a_{1}...a_{N}\underset{def}{=}\mathbf{%
[}n-k]_{i-1}^{N}G^{(k)}x_{1}...x_{n-(k+1)}\underset{n-k}{a_{i}}a_{1}...a_{N}$%
, for $k+1<n$

\noindent $G^{(n)}a_{1}...a_{N}\underset{def}{=}[1]_{i=1}^{N}G^{(n-1)}%
\underset{1}{a_{i}}a_{1}...a_{N}.$

Evidently $G^{(1)}$ is totally symmetric in the $a_{k}$'s, and if $G^{(k)}$
is, so is $G^{(k+1)}$; hence, by induction on $k$, so is $G^{(n-1)}$;
whereupon so also is $G^{(n)}.$ The logical equivalences $A\vDash
T\leftrightarrow T^{(1)}$, $A\vDash T^{(k)}\leftrightarrow T^{(k+1)}$ are
obvious, hence, again by induction on $k$, $A\vDash T\leftrightarrow T^{n}$.
Defining $T_{2}$ as $T^{(n)}\wedge A$, (iii) follows.

Now define $T_{S}$ as the $L$-sentence obtained by replacing every
occurrence of $a_{k}$ in $T_{2}$ (i.e. in $T^{(n)}\wedge A$) by $x_{k}$, $%
k=1,..,N$, and preface the expression that results by $N$ existential
quantifiers. Obtain in this way:

$\exists x_{1}...\exists x_{N}\left( G^{(n)}x_{1}...x_{N}\wedge
\dbigwedge\limits_{i,j=1,i\neq j}^{N}x_{i}\neq x_{j}\wedge \forall
x\dbigvee_{k=1}^{N}x=x_{k}\right) \underset{def}{=}\exists x_{1}...\exists
x_{N}Gx_{1}...x_{N}$

Then (v) is immediate. Since $G^{(n)}$ is totally symmetric, so is $G$, as
required. It only remains to prove (ii). But in any model in which $A$ is
true and $T_{S}$ is true, $Ga_{\pi (1)}...a_{\pi (N)}$ is true for some
choice of permutation $\pi .$ Since $G$ is totally symmetric, $T_{2}$ is
true as well\medskip
\end{proof}

\medskip \bigskip 

\noindent {\Large References\bigskip }

\noindent Albert, D. (2000). \textit{Physics and Chance}. Cambridge: Harvard
University Press.

\noindent Bach, A. (1990). Boltzmann's probability distribution of 1877. 
\textit{Archive for the History of the Exact Sciences}, \textit{41,} 1-40.

\noindent Bach, A. (1997). \textit{Indistinguishable Classical Particles}.
Berlin: Springer.

\noindent Costantini, D. (1987). Symmetry and the indistinguishability of
classical particles. \textit{Physics Letters A}, \textit{123}, 433-6.

\noindent Dieks, D. (1990). Quantum statistics, identical particles and
correlations. \textit{Synthese}, \textit{82}, 127-55.

\noindent Einstein, A. (1928) General discussion. \textit{\'{E}lectrons et
photons : Rapports et discussions du cinqui\`{e}me Conseil de physique tenu 
\`{a} Bruxelles du 24 au 29 octobre 1927 sous les auspices de l'Institut
international de physique Solvay. }Paris: Gauthier-Villars.

\noindent Ehrenfest, P. (1911). Welche Z\"{u}ge der Lichquantenhypothese
spielen in der Theorie der W\"{a}rmestrahlung eine wesentliche Rolle? 
\textit{Annalen der Physik,} \textit{36}, 91-118. Reprinted in Bush, (ed.), 
\textit{P. Ehrenfest, Collected Scientific Papers}. Amsterdam:
North-Holland, 1959.

\noindent Ehrenfest, P., and V. Trkal (1920). Deduction of the dissociation
equilibrium from the theory of quanta and a calculation of the chemical
constant based on this. \textit{Proceedings of the Amsterdam Academy}, 
\textit{23}, 162-183. Reprinted in P. Bush, (ed.), \textit{P. Ehrenfest,
Collected Scientific Papers}. Amsterdam: North-Holland, 1959.

\noindent van Fraassen, B. (1991). \textit{Quantum Mechanics:\ An Empiricist
View}. Oxford: Clarendon Press.

\noindent Gibbs, , J. W. \ (1902). \textit{Elementary Principles in
Statistical Mechanics}. New Haven: Yale University Press.

\noindent Gottesman, D. (2005), Quantum statistics with classical particles.
http://xxx.lanl.gov/cond-mat/0511207

\noindent Hestines, D. (1970). Entropy and indistinguishability. \textit{%
American Journal of Physics}, \textit{38}\textbf{,} 840-845.

\noindent Huang, K. (1963). \textit{Statistical Mechanics}. New York: Wiley.

\noindent Huggett, N. (1999). Atomic metaphysics.\textit{\ Journal of
Philosophy, 96}, 5-24.

\noindent van Kampen, N. (1984). The Gibbs paradox. \textit{Essays in
theoretical physics in honour of Dirk ter Haar,} W.E. Parry, (ed.). Oxford:
Pergamon Press.

\noindent Leinaas, J., and J. Myrheim (1977). On the theory of identical
particles, \textit{Il Nuovo Cimento}, 37B, 1-23.

\noindent Lieb, E., and J. Yngvason, (1999). The physics and mathematics of
the second law of thermodynamics.\textit{\ Physics Reports,} \textit{310},
1-96. Available online at arXiv:cond-mat/9708200 v2.

\noindent Natanson, L. (1911). \"{U}ber die statistische Theorie der
Strahlung. \textit{Physikalische Zeitschrift}, \textit{112}, 659-66.

\noindent Pauli, W. (1946). The exclusion principle and quantum mechanics.
Nobel Lecture, Stockholm, Dec. 13, 1946. Reprinted in C. Enz, K. von Meyenn
(eds.), \textit{Wolfgang Pauli: Writings on Physics and Philosophy}. Berlin:
Springer-Verlag, 1994.

\noindent Planck, M. (1900). Zur Theorie des Gesetzes der Energieverteilung
im Normalspectrum. \textit{Verhandlungen der Deutsche Physicakalishe Gesetzen%
}, \textit{2}, 202-204. Translated in D. ter Haar (ed.), \textit{The Old
Quantum Theory}. Oxford: Pergamon Press, 1967.

\noindent Planck, M. (1912). La loi du rayonnement noir et l'hypoth\`{e}se
des quantit\'{e}s \'{e}l\'{e}mentaires d'action. In P. Langevin and M. de
Broglie (eds.), \textit{La Th\'{e}orie du Rayonnement et les Quanta -
Rapports et Discussions de la R\'{e}sunion Tenue \`{a} Bruxelles, 1911}.
Paris: Gauthier-Villars.

\noindent Planck, M. (1921). \textit{Theorie der W\"{a}rmesrahlung. }4th
ed., Leipzig: Barth, Leipzig.

\noindent Pniower, J. (2006). \textit{Particles, Objects, and Physics}. D.
Phil Thesis, Oxford: University of Oxford.

\noindent Poincar\'{e}, H. (1911). Sur la theorie des quanta. \textit{%
Comptes Rendues,} \textit{153}, 1103-1108.

\noindent Poincar\'{e}, H. (1912). Sur la theorie des quanta. \textit{%
Journal de Physique,} \textit{2,} 1-34.

\noindent Pooley, O. (2006). Points, particles, and structural realism. \ In
D. Rickles, S. French, and J. Saatsi (eds.),\textit{The Structural
Foundations of Quantum Gravity}. Oxford: Clarendon Press.

\noindent Rapp, D. (1972). \textit{Statistical Mechanics}. New York: Holt,
Rinehart and Winston.

\noindent Rosenfeld, L. (1959). Max Planck et la definition statistique de
l'entropie. \textit{Max-Planck Festschrift 1958}, Berlin: Deutsche Verlag
der Wissenschaften. Trans. as Max Planck and the Statistical Definition of
Entropy. In R. Cohen and J. Stachel (eds.), \textit{Selected Papers of Leon
Rosenfeld}. Dordrecht: Reidel, 1979.

\noindent Saunders, S. (2003a). Physics and Leibniz's principles. In K.
Brading and E. Castellani (eds.), \textit{Symmetries in Physics: New
Reflections}. Cambridge: Cambridge University Press.

\noindent Saunders, S. (2003b). Indiscernibles, general covariance, and
other symmetries: the case for non-reductive relationalism. In A. Ashtekar,
D. Howard, J. Renn, S. Sarkar, and A. Shimony (eds.), \textit{Revisiting the
Foundations of Relativistic Physics: Festschrift in Honour of John Stachel}.
Amsterdam: Kluwer.

\noindent Saunders, S. (2006). Are quantum particles objects? \textit{%
Analysis}, forthcoming.

\noindent Schr\"{o}dinger E. (1946). \textit{Statistical Thermodynamics}.
Cambridge: Cambridge University Press.

\noindent Shankar, R. (1980). \textit{Principles of Quantum Mechanics}. New
Haven: Yale University Press.

\noindent Stern, O. (1949). On the term $N!$ in the entropy. \textit{Reviews
of Modern Physics}, \textit{21}, 534-35.

\noindent Stachel, J. (2002). The relations between things' versus `the
things between relations':the deeper meaning of the hole argument. In D.
Malement (ed.), \textit{Reading Natural Philosophy/Essays in the History and
Philosophy of Science and Mathematics}. Chicago and LaSalle: Open Court.

\bigskip 

\bigskip 

\bigskip 

\bigskip 

\noindent

\noindent

\noindent

\noindent

\noindent

\end{document}